\documentclass{ws-mpla}
\usepackage[super]{cite}
\usepackage{graphicx}
\begin{document}

\markboth{Steven D. Bass}{Cosmology with an emergent Standard Model}

\catchline{}{}{}{}{}

\title{Cosmology with an emergent Standard Model\\
}

\author{Steven D. Bass
}

\address{
Kitzb\"uhel Centre for Physics, Kitzb\"uhel, Austria \\
Institute of Physics, 
Jagiellonian University in Krak\' ow, Krak\'ow, Poland \\
Steven.Bass@cern.ch
}



\maketitle

\pub{Received (Day Month Year)}{Revised (Day Month Year)}

\begin{abstract}
We discuss new ideas that the Standard Model might be emergent
with connection to electroweak vacuum stability 
and related consequences for cosmology.
In this scenario, 
the gauge symmetries and particles of the Standard Model would be ``born'' in some phase transition at a large scale about $10^{16}$~GeV with the Standard Model parameters constrained by the requirement of vacuum stability.
Emergent gauge symmetries are well known in condensed matter physics. Perhaps the Standard Model might also be emergent.
In this case the particles would be the stable long range excitations of degrees of freedom that operate above the scale of emergence.
The dark energy scale comes out with similar size to the tiny masses of light Majorana neutrinos. The emergence scenario comes with interesting constraints on possible dark matter structure.
New physics at energy scales around $10^{16}$~GeV 
might be explored through its effects in the neutrino mass matrix plus using high frequency gravitational waves and polarisation observables in the cosmic microwave background.
%

\end{abstract}


\section{Introduction}

Particle physics and cosmology are inspired by the quest to go beyond the Standard Model and General Relativity in our understanding of Nature. 
A key question is the origin of the gauge symmetries and degrees of freedom that determine particle physics dynamics \cite{wspcbook}.

The Standard Model~\cite{Pokorski:2000ruo,Taylor:1976ru,Altarelli:2013tya}, SM, 
is working exceptionally well in present particle physics experiments, 
from LHC high-energy collider energies through to  
low-energy precision measurements. 
There is no evidence so far in the data for new particles or interactions. 
The SM is a quantum field theory built on the gauge groups of U(1)$_{\rm Y} \otimes$ SU(2)$_{\rm L} \otimes$ SU(3)$_{\rm c}$ 
with the gauge bosons being the massless photon of QED and gluons of QCD plus the massive W and Z that mediate the weak interactions.
The discovery of the Higgs boson at CERN in 2012  completes the particle spectrum of the SM \cite{Bass:2021acr,Jakobs:2023fxh}.
In spite of this success we know that the SM is not the whole story~\cite{Pokorski:2023ukp}.
Open puzzles include the origin of  neutrino masses, 
the matter--antimatter asymmetry in the Universe and 
issues of dark energy, DE, and dark matter, DM, 
plus primordial inflation at the interface of particle physics with gravitation. 
One also wants to understand the origin of the gauge symmetries and the fermion families, viz. why are there three? 
How high in energy might the SM apply without the need for new interactions and at what energies might the new physics enter?
In seeking to go beyond the SM 
(and General Relativity, GR), should we be looking for new particles and/or new principles?

Experiments are pushing the frontiers looking for cracks 
in our description of Nature and searching for clues to new physics.
These experiments involve each of high-energy colliders, 
low-energy precision measurements,  
astroparticle physics,  gravitational waves and cosmology surveys. 
There is also the deeper theoretical question how to combine the SM and GR. 
GR is a classical theory. 
Should gravity even be quantised?

DE and DM are at the particle-gravity frontier.
DE drives the accelerating expansion of the Universe. 
Cosmology also requires some extra non-luminous DM not made of baryonic matter.
DM clumps like normal SM matter under gravitation whereas the DE (or cosmological constant) is the same everywhere.
So far these are perceived only through their gravitational interactions as measured in astrophysics and cosmology surveys.
The energy budget of the Universe has 68\% in DE, 5\% in SM matter and 27\% in non-luminous DM with structure to be determined.
The Universe is spatially flat on 
distance scales 
greater than galactic sizes.
There is a vigorous program of experiment and theory trying to understand the origin and possible dynamics of this ``dark sector''.
The DE measured in astrophysics is characterised by a vacuum energy density scale 0.002 eV \cite{Planck:2018vyg}.
This value is very much smaller than SM scales from QCD and electroweak interactions  
despite the contribution of SM quantum contributions
(zero-point energies and potentials) in the vacuum.
The SM scales are themselves very much less than the 
Planck scale despite quantum corrections naively pushing the Higgs mass into the ultraviolet and naturalness arguments associated with treating the SM as a low-energy effective theory~\cite{Wells:2009kq}.
Astrophysics observations tell us that there is a need for dark matter.  
So far, in terms of experimental guidance, we do not know even the starting point in unraveling its internal structure.
Ideas range from possible  ultralight new particles through to black holes and modifications of usual gravity theory.

In the absence of new physics, it makes sense to extrapolate the SM to the highest scales and to look for consistency issues.
LHC data while so far not revealing any evidence for new particles or interactions does come with the fascinating issue of vacuum stability with the Higgs self-coupling $\lambda$ staying positive under renormalisation group evolution up to at least $10^{10}$ GeV and maybe up to the Planck scale if we assume just the SM with its measured masses and couplings and no coupling to new particles or interactions \cite{Bednyakov:2015sca,Jegerlehner:2013cta,Degrassi:2012ry,Buttazzo:2013uya,Alekhin:2012py,
Masina:2012tz,Hiller:2024zjp}.
The Higgs couplings to the SM particles 
W, Z, t, b, $\tau, \mu$ are presently known to 10\% accuracy and are in good agreement with the SM. 
The Higgs self-coupling awaits accurate measurement.
With high-energy observables so far in excellent agreement with the SM, the SM relation 
$\lambda = m_{\rm h}^2/2 v^2$ relating $\lambda$ to the Higgs mass $m_{\rm h}$ and vacuum expectation value $v$ is presently assumed.
The high luminosity upgrade of the LHC 
should measure $\lambda$ to $\approx 28\%$ precision  \cite{CMS:2025hfp}. 
Going further, 
a 5\% accurate measurement could be made with a future 100 TeV centre-of-mass proton-proton collider~\cite{Jakobs:2023fxh}.
Taking the measured SM inputs, renormalisation group evolution 
gives that 
$\lambda$ and its
$\beta-$function should vanish deep in the ultraviolet.
If $\lambda$ crosses zero below the Planck scale $M_{\rm Pl}$, this signifies a metastable vacuum if the SM is extrapolated up to $M_{\rm Pl}$.
One finds that the 
SM Higgs vacuum comes out very close to the border of stable and metastable, within about one standard deviation of being stable up to the $M_{\rm Pl}$~\cite{Bednyakov:2015sca} and with the most metastable solution having a life time $\sim 10^{600}$ years~\cite{Buttazzo:2013uya}, much greater than the present age of the Universe. 
Modulo the large renormalisation group extrapolation involved, 
this result may be hinting at 
some possible new critical phenomena in the ultraviolet \cite{Degrassi:2012ry,Jegerlehner:2013cta,Buttazzo:2013uya}. 
The vacuum stability is very sensitive to exact values of SM parameters, 
especially the Higgs and top quark masses and the QCD coupling $\alpha_s$. 
That is, it involves a delicate fine tuning and conspiracy of SM parameters.
Small changes in these parameters can lead to very different physics.
\footnote{Small changes in the values of SM parameters 
like the up and down quark masses and the fine structure constant 
can also violate the conditions needed 
for, e.g.,  
big bang nucleosynthesis and the conditions needed to support life as we know it.}
If SM vacuum stability is taken as a guiding principle, then the  
SM parameters may be correlated with physics deep in the ultraviolet with an implicit reduction in the number of fundamental couplings.

The SM when taken alone then comes with three key scales: the QCD and electroweak scales, 
$\Lambda_{\rm QCD} \approx 300$ MeV and
$\Lambda_{\rm ew} \approx 246$ GeV respectively, 
plus a large scale in the ultraviolet
where the Higgs self-coupling crosses zero 
(assuming it does below the Planck scale)
which is linked to the stability of the  vacuum and which might be taken as the scale of ultraviolet completion if we require the vacuum to be fully stable.
Independent of this physics, 
considerations based on tiny neutrino masses, dark energy and inflation suggest that interesting new dynamics might enter around 
$\sim 10^{16}$ GeV. 
Perhaps these scales might be related.

Inspired by the idea of possible new critical phenomena in the ultraviolet, 
this article explores the paradigm that the particles and local
gauge symmetries of the SM might be emergent below some large scale about $10^{16}$ GeV, 
``born'' in a phase transition deep in the ultraviolet and connected with the stability of the electroweak vacuum.
Emergent gauge symmetries are well known in condensed matter physics. 
Might the gauge symmetries of the SM also be emergent?
With the SM working so well in our experiments, 
emergence offers  plausible answers to some of the key open puzzles in particle physics and its interface with cosmology. 
This Brief Review gives an introduction to the key ideas and connection to experiments.
More detailed discussion is found in the  book~\cite{wspcbook}.

\section{An emergent Standard Model}

An important issue is the origin of the gauge symmetries that determine the interactions of the SM. 
Where might they come from?
Gauge symmetries can be emergent, appearing only below some (very) large scale in the ultraviolet. 
Emergent gauge symmetry ideas for particle physics have been developed in Refs.~\cite{Jegerlehner:2013cta,Jegerlehner:1978nk,Jegerlehner:1998kt,Jegerlehner:2018zxm,Jegerlehner:2021vqz,Bass:2020egf,Bass:2021wxv,Bass:2023ece,wspcbook,Bjorken:1963vg,Bjorken:2001pe,Forster:1980dg,Chkareuli:2001xe,tHooft:2007nis,Witten:2017hdv,Wilczek:1984dh,Wetterich:2016qee,Barcelo:2016xhp,Barcelo:2021idt,Halimeh:2021vzf}.

\subsection{Emergent gauge symmetries in condensed matter systems}

What do we mean by emergence? 
Emergence in physics occurs when a many-body system exhibits collective behaviour in the infrared that is qualitatively different from that of its more primordial constituents as probed in the ultraviolet \cite{Anderson:1972pca,Palacios:book}.
For example, 
classical physics is emergent from quantum physics. 
Hadrons like protons, neutrons and pions are emergent from quark-gluon degrees of freedom. 
Chemistry and biology are emergent from electrodynamics.
An interesting case of emergent phenomena from everyday 
experience is the collective change in the travel direction of starling flocks from individual bird's flight fluctuations.
Symmetries can also be emergent. 
As an
everyday example of emergent symmetry, consider a carpet which looks flat and translational
invariant when looked at from a distance. Up close, e.g., as perceived by an ant crawling on it,
the carpet has structure and this translational invariance is lost. The symmetry perceived in the infrared, e.g., by someone looking at it from a distance, ``dissolves'' in the ultraviolet when the
carpet is observed close up.

Examples of emergent gauge phenomena in condensed matter physics 
include high temperature superconductors~\cite{Baskaran:1987my,Affleck:1988zz,Sachdev:2015slk}, 
 superfluid $^3$He-A \cite{Volovik:2003fe,Volovik:2008dd},
string-nets~\cite{Wen:2007joe,Levin:2004js}, 
spin ice phenomena in magnetic systems~\cite{Rehn:2016eqc} and 
the quantum Hall effect \cite{Tong:2016kpv} 
plus various more simple quantum systems~\cite{Wilczek:1984dh}.
Here the emergent gauge symmetries are associated with  
so-called topological phase transitions and with long range quantum entanglement.
For extra review type discussions, see Refs.~\cite{Powell:2020osu,Zaanen:2011hm,Moessner:book}.
Emergent gauge fields play an essential role in quantum simulations of quantum field theories~\cite{Banerjee:2012pg,Zohar:2015hwa}.
Topological phase transitions occur without local order parameters, in contrast to usual Ginzburg-Landau phase transitions which do. 
Collective gauge fields beyond the more fundamental photons of QED 
can ``emerge'' from the quantum structure of the many-body ground state.  
The ground state can exhibit multiple degeneracy with the degenerate 
substates being related by emergent gauge transformations. 

The prototype condensed matter system is the Fermi-Hubbard model of 
strongly correlated electrons in a two dimensional atomic lattice at 
half-filling.  
In the low-energy limit the system exhibits SU(2) and U(1) emergent gauge symmetries and spin-charge separation~\cite{Baskaran:1987my,Affleck:1988zz}. 
The emergent SU(2) couples to the spin of the electrons which then becomes dynamical to internal observers.
This model is a starting point for many discussions of 
high temperature  superconductors. 
Any additional fluctuations of the atomic lattice sites corresponds to extra bosonic phonon excitations beyond the emergent gauged system independent of whether the lattice ions are fermions or bosons.

A further interesting example is the A-phase of superfluid $^3$He 
where the topology in momentum space plays an essential role 
\cite{Volovik:2003fe,Volovik:2008dd}.
With superfluid $^3$He the Cooper pairs are in $p-$wave.
In the A-phase they have just $S_z =\pm 1$ states with the $S_z=0$ state absent.
This phase forms at 2.6 mK and 21 bars of pressure and differs from the so called B-phase where the Cooper pairs have total angular momentum $J=0$.
The A-phase exhibits Fermi points -- singular points in momentum space where the fermion quasiparticles have no mass gap.
Close to the Fermi points the quasiparticles of $^3$He-A are chiral fermions plus emergent
SU(2) and U(1) spin-one gauge bosons and spin-two effective ``gravitons''~\cite{Volovik:2003fe,Volovik:2008dd}. 
One finds emergent local gauge interactions with spin of the Helium quasiparticles becoming dynamical to internal observers. 
The gauge symmetries correspond to the freedom 
(degeneracy) in choosing the position of the Fermi point on the former Fermi surface.
The quasiparticles in $^3$He-A, both fermions and gauge bosons, 
each come with a common limiting velocity behaving like a relativistic quantum field theory like what happens with Lorentz invariance in the SM (for details see Sect. 9.3.2 of Ref.~\cite{Volovik:2003fe}).

\subsection{Towards an emergent SM}

How might the particle physics SM be emergent?
There are ideas connected to phase transitions, to renormalisation group decoupling and to spontaneously broken Lorentz symmetry, SBLS.

First, a statistical 
system near its critical point has the interesting feature that its
long range asymptote behaves as a quantum field theory with properties described by the renormalisation group~\cite{Wilson:1973jj,Peskin:1995ev}.
The famous example connects Ising like systems and $\phi^4$ theory with the critical dimension equal to four.
If the spectrum includes $J=1$ excitations among 
the degrees of freedom in the low-energy phase, then it is a gauge theory. 
(Renormalisable quantum field theories with $J=1$ vector particles 
exhibit local gauge symmetries
\cite{tHooft:1979hnm}.) %
Gauge symmetries would then be an emergent property of the low-energy phase and ``dissolve'' in a phase transition deep in the ultraviolet 
~\cite{Jegerlehner:1978nk,Jegerlehner:1998kt,Jegerlehner:2013cta,Forster:1980dg,Bjorken:2001pe,tHooft:2007nis,
Bass:2021wxv,Bass:2023ece}.
The quarks and leptons as well as the gauge bosons and
Higgs boson might be the stable collective long-range excitations of some (unknown) more primordial degrees of freedom that exist above the scale of emergence.
The vacuum of the low-energy phase should be stable below the scale of emergence.
In this case, the SM might saturate the particle spectrum at mass dimension $D=4$ with a stable vacuum corresponding to positive Higgs 
self-coupling $\lambda$ below the scale of emergence.
The Higgs mass would then be environmentally selected in connection with vacuum stability and the phase transition that produces the SM.

Suppose the SM works like this.
Then it behaves as an effective theory with the renormalisable 
theory at $D=4$ supplemented by a tower of non-renormalisable higher dimensional operators each 
suppressed by powers of the large scale of emergence~\cite{Jegerlehner:2013cta,Witten:2017hdv}. 
The global symmetries of the SM at $D=4$ 
are constrained by gauge invariance and renormalisability.
The higher dimensional operators are less constrained 
and may exhibit extra global symmetry breaking. 
Lepton number violation and tiny Majorana neutrino masses 
may enter at $D=5$, 
meaning that they are suppressed by a single power of
the scale of emergence through the so-called Weinberg operator 
\cite{Weinberg:1979sa}. 
One finds neutrino masses 
$m_\nu \sim \Lambda_{\rm ew}^2/M$ 
where 
$\Lambda_{\rm ew}$ 
is the electroweak scale and $M$ is the scale of emergence. 
New CP violation, needed for baryogenesis, might occur in 
the Majorana phases at $D=5$ and also in new $D=6$ operators 
\cite{Grzadkowski:2010es}, meaning that they are suppressed by two powers of $M$.
Proton decays might occur at $D=6$ 
\cite{Weinberg:1979sa,Wilczek:1979hc}.
If present, new pseudoscalar axion particles 
come with masses and couplings to SM particles entering at $D=5$~\cite{Weinberg:1977ma,Wilczek:1977pj}. 
Lorentz symmetry violations might occur at $D=6$~\cite{Bjorken:2001pe}.
For extra discussion of the SM as an effective theory including higher dimensional operators, see Ref.~\cite{Weinberg:2018apv}.

Within this emergence picture,  
if one increases the energy much above the
electroweak scale, then the physics becomes increasingly  symmetric with energies 
$E \gg \Lambda_{\rm ew}$ 
until we come within about 0.1\% or so of the scale of emergence. 
Then new global symmetry violations in the higher dimensional operators become important so the physics becomes increasingly chaotic until one goes through the 
phase transition associated with the scale of emergence,
with the physics above this scale 
then described by new degrees of freedom and perhaps new physical laws.
Any perturbative extrapolation of SM degrees of freedom above the scale of emergence would reach into an unphysical region.
This scenario contrasts with unification models which involve maximum symmetry in the extreme ultraviolet. 
In unification models 
the gauge couplings of the SM would meet at some large scale. 
If one takes the minimal SM with no extra interactions, then the gauge couplings almost meet but not quite.
LHC data (so far) reveal no evidence for higher dimensional correlations in searches for new operator terms 
divided by powers of a large mass scale below the few TeV range \cite{Slade:2019bjo,Ellis:2020unq}.
The tiny neutrino masses indicated by neutrino oscillation experiments as well as constraints on proton decays and Lorentz invariance suggest a scale of emergence deep in the 
ultraviolet, 
much above the Higgs and other SM particle masses.

Going further, 
keeping with the idea of maximum emergent symmetry in the infrared,
it is natural to expect the SM vacuum to be global spacetime translation invariant in the leading term at $D=4$.
On the other hand, global spacetime translational invariance
is violated by a finite cosmological constant which acts like the vacuum energy density perceived by gravitation and as a gravitational source which curves the spacetime, thus violating the global spacetime translational invariance of the vacuum.
Any violation might come with scale  suppressed by a factor of $M$.
This idea underpins emergence ideas about the cosmological constant and DE 
\cite{Bass:2020egf,Bass:2023ece,Bass:2020nrg},
see Section 3.2.

In thinking about the Higgs mass hierarchy puzzle with electroweak symmetry breaking and Higgs vev 
$v \ll M_{\rm Pl}$, 
it is interesting 
to consider  
possible similar phenomena with the ferromagnetic 
phase transition in condensed matter physics. 
Below the phase transition the magnetisation is very small 
close to the phase transition when we approach the critical temperature $T_c$ from below, viz. 
when the reduced temperature
$(T-T_c)/T \to 0^-$. 
Whereas emergent gauge systems can be associated with topological like
phase transitions without a local order parameter, 
Higgs phenomena is associated with spontaneous symmetry breaking defined with respect to a particular gauge choice~\cite{Kibble:2014gug} with all choices of gauge being equivalent.
Electroweak symmetry breaking might correspond to a Universe 
close to the phase transition and very near to the critical point \cite{Jegerlehner:2013cta}, as also hinted at with electroweak vacuum stability.
With a possible emergent SM, there is a fundamental difference
with emergent condensed matter systems in that there we know the degrees of freedom above and below the phase transition whereas with an emergent particle physics we measure only the low energy phase in our experiments.

Besides through possible phase transitions discussed above, 
emergent gauge symmetries can also appear through the decoupling of gauge violating terms 
in the renormalisation group at an infrared fixed point 
(where couplings become invariant under changes in the scale or resolution)
\cite{Wetterich:2016qee}
and also in connection with possible spontaneous breaking of Lorentz symmetry~\cite{Bjorken:1963vg,
Bjorken:2001pe,Chkareuli:2001xe}.
In the former case, 
the coefficient of any local gauge symmetry violating term blows up at the fixed point, 
in contrast to restoration of global symmetries where the coefficient of any symmetry violating term goes to zero at the fixed point. 
With SBLS, 
non observability of any Lorentz violating terms at $D=4$ 
corresponds to gauge symmetries for vector fields like the photon. 
Possible Lorentz violation here might be manifest in terms at largest with size ${\cal O} (\Lambda_{\rm ew}^2/M^2)$ 
with a preferred reference frame naturally identified with 
the frame where the cosmic microwave background, CMB, 
is locally at rest \cite{Bjorken:2001pe}.

\section{Key observables}

Suppose that the SM might be emergent. 
How might this be manifest in particle physics 
phenomenology and in cosmology?
In this Section we give a brief overview of some of the main 
consequences for unsolved puzzles of beyond the Standard Model particle physics. 
Phenomenologically, we will find that an interesting value for the scale of emergence is $M \sim 10^{16}$ GeV, 
which is within the range where $\lambda$ might cross zero under perturbative renormalisation group evolution.

\subsection{Neutrinos}

Majorana neutrinos enter with masses given though the $D=5$ 
Weinberg operator 
\begin{equation}
    m_\nu \sim \Lambda_{\rm ew}^2/M
    \label{eq1}
\end{equation}
and accompanying lepton number violation with signature to be looked for in neutrinoless double $\beta-$decay experiments. 
With Majorana neutrinos there are 3 new CP phases.
Neutrino oscillation experiments, where neutrinos created with a particular flavour are later measured to have a different flavour, 
point to the existence of tiny neutrino masses.
Assuming three species of neutrinos, the neutrino oscillation data 
constrains the largest mass squared difference to be
$\approx 2 \times 10^{-3}$ eV$^2$ 
with the smaller one as 
$(7.53 \pm 0.18) \times 10^{-5}$ eV$^2$
\ 
\cite{BahaBalantekin:2018ppj}.
With these values the lightest neutrino mass is expected to be about 10$^{-8}$ times the value of the electron mass corresponding to 
$M \sim 10^{16}$ GeV in Eq.~(\ref{eq1}).

\subsection{Dark energy}

Dark energy drives the  accelerating expansion of the Universe~\cite{Weinberg:1988cp,Peebles:2002gy}. 
The simplest explanation is a time independent 
cosmological constant $\Lambda$ 
in Einstein's equations of General Relativity 
\begin{equation}
R_{\mu \nu} - \frac{1}{2} g_{\mu \nu} \ R = 
- \frac{8 \pi G}{c^4} T_{\mu \nu} + \Lambda g_{\mu \nu} .
\label{eq2}
\end{equation}
Here $R_{\mu \nu}$ is the Ricci tensor, 
$R$ is the Ricci scalar
and 
$T_{\mu \nu}$ is the energy-momentum tensor 
for excitations above the vacuum;
$G$ is Newton's constant and $c$ is the speed of light. 
The cosmological constant comes proportional to the metric tensor $g_{\mu \nu}$
and with vacuum equation of state 
(energy density = - pressure).
Ideas about possible time dependence in the DE and its equation of state, EoS, are discussed in Section~\ref{sect:de} below.

The cosmological constant 
measures the vacuum energy density perceived by gravitation
\begin{equation}
\rho_{\rm vac} = \Lambda 
\times 
 c^4 
/ (8 \pi G ) 
\label{eq3}
\end{equation}
with an associated scale $\mu_{\rm vac}$,
$\rho_{\rm vac} = \mu_{\rm vac}^4$.
Astrophysics observations \cite{Planck:2018vyg} 
tell us 
that $\Lambda = 1.088 \times 10^{-56} \ {\rm cm}^{-2}$ corresponding to a cosmological constant scale 
(in natural units)
\begin{equation}
\mu_{\rm vac} = 0.002 {\rm \ eV} .
\label{eq4}
\end{equation}
The present period of accelerating expansion is observed in Supernova 1a data to have began about 5 billion years ago when the matter density of the expanding Universe fell below 
$\rho_{\rm vac}$, which then took over as the main  driving term for the expansion.
The Universe has a  spatially flat geometry today on large distance scales in FLRW cosmology.

The small scale 0.002 eV in Eq.~(\ref{eq4}) is very small 
compared to usual particle physics scales that characterise the 
SM particle physics vacuum.
A priori, 
one expects the vacuum energy to be sensitive to quantum fluctuations 
(zero-point energies, ZPEs) 
and potentials in the vacuum with these terms involving the much larger QCD and electroweak  scales,
viz.  
\begin{equation}
    \mu_{\rm vac} \ll \Lambda_{\rm ew}, m_{\rm h} \ll M_{\rm Pl} .
    \label{eq5}
\end{equation} 
Here $M_{\rm Pl}$ is the Planck scale where quantum gravity effects might become important. 
One also finds an extra ``bare gravitational contribution'' 
$\rho_\Lambda$ 
to the net $\rho_{\rm vac}$.~\footnote{
See Ref.~\cite{Bass:2023ece} for a review of these scale hierarchies plus the  sensitivity of individual ZPE, potential and gravitational contributions to the symmetry properties of the ultraviolet  regularisation in their calculation.}
Why is the net cosmological constant so small?

The cosmological constant is connected with the symmetries of the metric $g_{\mu \nu}$. 
With a finite cosmological constant Einstein's equations of gravitation have no vacuum solution where $g_{\mu \nu}$ is the constant Minkowski metric.
That is, 
global spacetime translational invariance of the vacuum is broken by a finite value of $\Lambda$
\cite{Weinberg:1988cp}.
The reason is that a non-zero cosmological constant acts as a gravitational source which generates a dynamical spacetime  
 with accelerating expansion of the Universe for positive $\Lambda$.
Suppose that the vacuum including condensates with finite vacuum expectation values is spacetime translational invariant and that  
flat spacetime is consistent at mass  dimension four, 
just as suggested by the success of the SM.
With the SM as an effective theory emerging in the infrared, 
the low-energy global symmetries including spacetime translation invariance can be broken through additional higher dimensional terms 
suppressed by powers of the large scale of emergence $M$.
QCD and electroweak interactions are characterised by the scales  
$\Lambda_{\rm QCD} \approx 300$ MeV and 
$\Lambda_{\rm ew} \approx 246$ GeV. 
These scales might then enter the cosmological constant
with the scale of the leading term suppressed by the factor  
$\Lambda_{\rm ew}/M$ 
 -- see Refs.~\cite{Bass:2020nrg,Bass:2020egf} and the early work
\cite{Bjorken:2001pe,Bjorken:2001yv}.
This scenario, if manifest in nature, 
would explain why the cosmological constant scale $\mu_{\rm vac} = 0.002$ eV 
is similar to what we expect for light neutrino masses  
\cite{Altarelli:2004cp} assuming that neutrinos are Majorana with masses given by the Weinberg operator~\cite{Weinberg:1979sa}.
One finds
\begin{equation}
 \mu_{\rm vac} \sim 
 m_\nu \sim \Lambda_{\rm ew}^2/M   .
 \label{eq6}
\end{equation}
In this scenario, the cosmological constant would vanish at mass dimension four.
This is equivalent to a renormalisation condition $\rho_{\rm vac} =0$
at $D=4$ 
imposed by global space-time translational invariance of the vacuum, 
even in the presence of QCD and Higgs condensates.
Taking the value 
$\mu_{\rm vac} = 0.002 {\rm \ eV}$
from astrophysics together with $\Lambda_{\rm ew} = 246$ GeV gives the value $M \sim 10^{16}$ GeV.

This scale $M \sim 10^{16}$ GeV 
is within the range where the Higgs 
self-coupling 
$\lambda$ might cross zero (if indeed it does) if the SM is extrapolated up to the highest energy scales using renormalisation group evolution.
It is also similar to the ``GUT scale" 
that typically  appears in unification models plus the number usually taken as the scale of inflation.
The observations of flatness, 
isotropy and homogeneity plus
the absence of magnetic monopoles 
which might have been produced in the very early Universe motivate primordial inflation ideas 
\cite{Baumann:2008bn} 
where initial exponential expansion is driven by an 
inflaton scalar field with slow-roll potential condition. 
Besides the present accelerating expansion, 
it is believed that the Universe experienced an initial period of exponential expansion called inflation, 
with factor at least $10^{26}$ or 60-e-folds expansion 
in the first about $10^{-33}$ seconds. 
In this picture 
the scale of inflation is typically taken as around 
$10^{16}$ GeV, 
which is close to our scale of emergence $M$.
Might the physics underlying inflation and SM emergence be connected?
If yes, the physics above the scale of inflation might involve 
new degrees of freedom and new physical laws.

In all, 
$\rho_{\rm vac}$ receives contributions from ZPEs, potentials in the vacuum and a net 
``bare gravitational'' term $\rho_\Lambda$.
The net scale is set by global spacetime translational invariance of the vacuum plus breaking by SM related terms in higher dimensional operators.
The ZPE and potential terms are cancelled by this $\rho_\Lambda$ 
contribution to preserve global 
spacetime translational invariance of the vacuum in the 
low-energy system characterised by the SM \cite{Bass:2023ece}.
A similar effect occurs in condensed matter physics with the Gibbs-Duhem relation for quantum liquids. 
Here the zero-point energy from low temperature  quasiparticles is cancelled by the effect of macroscopic degrees of freedom above the characteristic energy for the quantum liquid effective theory \cite{Volovik:2004gi}.
The cosmological constant need not jump through the QCD phase transition or crossover transition in the early Universe.
Any change in the ZPE and potential terms might be  compensated by changes in the gravitational term, with 
$\rho_\Lambda$ 
interpreted as parametrising the effect of physics above the scale of emergence.

\subsection{Time dependent dark energy}
\label{sect:de}

Looking beyond a simple time independent cosmological constant, 
searches for possible time dependence in 
DE are presently a topic vigorous investigation 
with new ongoing 
and planned cosmological surveys including the DESI~\cite{DESI:2025zgx}, LSST~\cite{LSSTScience:2009jmu} and Euclid~\cite{Amendola:2016saw} experiments.
Theoretical ideas include the vacuum expectation value of a possible 
time dependent extra scalar ``cosmon'' field
which interpolates between initial inflation and dark energy today~\cite{Wetterich:1994bg,Wetterich:1987fm,Peebles:1987ek,Peebles:2002gy} 
as well as running vacuum models 
\cite{SolaPeracaula:2022hpd}
and quantum breaking arguments~\cite{Dvali:2017eba}.

In the emergence scenario any time dependence might correspond to a change in 
$\Lambda_{\rm ew}^2/M$ 
(with $G$ and $c$ taken as fixed)
times its coefficient in the low-energy expansion. 
It is natural to want the Einstein equivalence principle 
to be working at $D=4$ 
in $\Lambda_{\rm ew}$ 
with any violation of the equivalence principle occurring in the higher dimensional term describing the cosmological constant. 
In this case,   
possible time dependence in the DE might occur in the scale of emergence or ultraviolet completion $M$ 
and/or in the coefficient of the
$\Lambda_{\rm ew}^2/M$ term that appears in the 
low-energy expansion. 
This time dependence would reflect  relaxation of the SM and
$\rho_{\rm vac}$
as one gets time-wise further away from the early Universe phase transition that produced it.
Condensed matter analogies are described in Refs.~\cite{Volovik:2004gi,Volovik:2006bh,Volovik:2023faj}. 
Possible Lorentz violation 
in higher dimensional operators~\cite{Bjorken:2001pe} 
might also induce some change in the DE EoS away from the vacuum solution $\rho_{\rm vac} = - p_{\rm vac}$.

\subsection{Dark matter}

Some extra stable and non-luminous dark matter without electromagnetic coupling is suggested by studies of properties of galaxies and galaxy clusters, gravitational lensing and the CMB
\cite{Baudis:2017avj,Bertone:2018krk,Wechsler:2018pic}.
The quest to understand this DM, 
$\approx 84\%$ 
of the total matter budget and
$\approx 27\%$ of the total energy budget of the Universe,
continues to inspire vast experimental and theoretical 
activity. Ideas include possible new types of elementary particles as well as black holes and possible modifications of usual gravity theory.
With new particles the mass scale is totally open with ideas including new particles with masses ranging from 
$10^{-22}$ eV up to 
$10^{15}$ GeV 
-- a range of 10$^{46}$. 
Discovery of the accelerating expansion of the Universe came through supernova type 1a observations 
\cite{SupernovaSearchTeam:1998fmf,SupernovaCosmologyProject:1998vns}.
Previously, 
a finite cosmological constant contribution was first suggested though
observation of a spatially flat Universe 
together with determinations of the dark matter contribution~\cite{Efstathiou:1990xe}.
Alternatively, with the cosmological constant scale deduced using the emergent Standard Model arguments above, 
Eq.~(\ref{eq6}), 
flatness gives a constraint on the amount of extra dark matter  needing extra theoretical understanding.

Suppose in the spirit of emergence that the SM saturates the physics at $D=4$. 
Small multiplets are preferred 
as collective excitations of the system that resides deep in the ultraviolet \cite{Jegerlehner:2013cta}.
In this case one does not expect lightest mass supersymmetry particles or inert extra Higgs doublet states as sources of dark matter. 
Possible DM candidates would include light mass pseudoscalar axions, 
BHs formed in the early Universe as well as relics/fluctuations  of the primordial system that might have frozen out along with the SM in the early Universe.

If present, 
axions come with masses and couplings to SM particles at $D=5$.
They are a non-leading term in the low-energy expansion in $1/M$. 
QCD axion masses are usually assumed from astrophysics and cosmology constraints to be between 
1 $\mu$eV and 3 meV. 
This mass range corresponds to an ultraviolet scale or ``axion decay constant'' 
between 
$6 \times 10^9$ and 
$6 \times 10^{12}$ GeV~\cite{Kawasaki:2013ae}.
While large, these scales are still less than the preferred 
emergence scale 
$M \sim 10^{16}$ GeV 
deduced in connection with the cosmological constant and neutrino masses.

One might also consider localised perturbations involving degrees of freedom that occur above the scale of emergence. 
Analogies with phenomena in condensed matter physics are discussed in  \cite{Klinkhamer:2016zzh,Jegerlehner:2023sfw}.
For example, 
perturbations in 
the physics associated with  $\rho_\Lambda$ 
might have occurred in the early Universe 
and couple to gravitation with a DM equation of state.
DM might perhaps behave similar to collective vibrations of the atomic ion lattice in superconductors.
Consider the extension of the Fermi-Hubbard model 
to include collective fluctuations of the atomic lattice sites.
The strongly correlated electron system described by the 
Fermi-Hubbard model
develops emergent gauge symmetries in the low-energy limit at half filling with the atomic lattice kept fixed. 
Vibrations of the atomic lattice system 
(phonons) 
exhibit bosonic statistics independent of the fermionic or bosonic nature of the atoms making up the lattice structure \cite{Moessner:book}. 
As a second example,
the main aspects of the BCS theory of 
superconductivity involve a two-fluid
model where the compressible electron gas and the related
phonon excitations 
propagate in an incompressible ion background. 
Two types of collective vibrational fields occur:  
the phonons that populate the long-range branch and the displacement field of the essentially incompressible ion system which could act as a dark matter like 
field~\cite{Jegerlehner:2023sfw}. 
Such ideas require detailed investigation to see whether they might have a chance to work.

\section{Conclusions and open questions}

The SM comes with the fascinating issue of vacuum stability which connects the world of experiments with the physics of the extreme ultraviolet.
To what maximum energies might the SM apply before new physics becomes important? 
In the absence of experimental evidence for new interactions it makes sense to take the SM and extrapolate it to the maximum energies and see what happens.
Vacuum stability is very sensitive to the exact values of SM parameters.
Criticality issues led us to consider the SM as a possible  emergent gauge system ``born'' in some topological like phase transition 
around $10^{16}$ GeV. In this case, new global symmetry breakings will occur in high dimensional operators with strength suppressed by powers of the large scale of emergence.
Examples include neutrino masses and accompanying CP phases plus possible axions at $D=5$ and new CP and baryon number violating terms at $D=6$.
There are interesting consequences for DE 
with the DE scale coming out similar to the Majorana mass scale at the same order in the low energy expansion in $1/M$.
There are also interesting consequences for possible DM scenarios.

For an emergent SM, an  
interesting theoretical  question involves the critical dimension for any phase transition that produces it. 
Might 3+1 dimensions be special? 
Also, what might be universality class of an emergent SM.
Might one find  parallels with condensed matter phenomena like the emergent gauge systems in 
$^3$He-A or string-nets?

How can we test these ideas and also explore 
possible new physics in the deep ultraviolet?
At collider energies we would like a precise measurement of the Higgs 
self-coupling at TeV scale energies to check that the SM is really working here.
With emergence we expect small multiplets meaning that 
SUSY, 2 Higgs doublet models... would be disfavoured.
New physics would enter in higher dimensional operators and  
at very high energies, about $10^{16}$ GeV, 
close to the scale of emergence.
Experiments with Majorana neutrinos
(assuming that neutrinos are Majorana as expected here) including CP phases might be sensitive to physics at these high scales.
Perhaps there are tiny effects with baryon number (proton decays) and Lorentz violation waiting to be found.
Also measurements of high frequency gravitational waves, HFGWs, with frequency about 1 GHz would be sensitive to 
possible phase transitions at these scales 
if these are first order and with a signal to be seen as a  
stochastic gravitational wave background~\cite{Domcke:2024soc,Aggarwal:2025noe}.
There are no known astrophysical sources at these frequencies. 
As an example of a condensed matter system with emergent gauge symmetries, 
the transition between the 
superfluid A and B phases of $^3$He 
(the A-phase with Fermi points and emergent U(1) and SU(2) gauge symmetries)
is first order whereas the transition to superfluidity itself is second order~\cite{Tian:2023,QUEST-DMC:2024crp}.
In particle physics 
if radiative corrections to the Higgs self-energy were to cross zero and change sign at a scale below the Planck scale
(a proposal which is calculation dependent and sensitive to the details of radiative corrections~\cite{Jegerlehner:2013cta,Degrassi:2012ry,Bass:2020nrg,Masina:2013wja,Hamada:2012bp}), 
this would result in a first order phase transition~\cite{Jegerlehner:2018zxm}.
Another key observable involves the tensor-to-scalar ratio and 
B-modes in CMB polarisation which are believed to be induced by gravitational waves propagating in the inflationary period~\cite{Baumann:2008bn,Komatsu:2022nvu}.
If the SM (and perhaps GR like the situation with $^3$He-A) 
might be emergent at a scale $\sim 10^{16}$ GeV, 
then the degrees of freedom above the scale of emergence and acting in the inflationary period  might be quite different. 

There are interesting challenges for experiments and theory.
If the emergence scenario really describes Nature,
then there are profound connections between the very big and very small waiting to be explored.

\section*{Acknowledgments}

I thank Fred Jegerlehner and
Stefan Pokorski for helpful discussions on the topics discussed here.

\end{document}